\documentclass[%
 aip,
rsi,%
 amsmath,amssymb,
 preprint,%
]{revtex4-1}

\usepackage{graphicx}
\usepackage{dcolumn}
\usepackage{bm}
\usepackage{color}
\usepackage{subfigure}
\usepackage{amsmath}
\usepackage[mathlines]{lineno}

\begin{document}

\preprint{}

\title{A Surface Impedance-Based Three-Channel Acoustic Metasurface Retroreflector}

\author{Chen Shen}
\affiliation{ 
Department of Electrical and Computer Engineering, Duke University, Durham, North Carolina 27708, USA
}%
\author{Ana~D\'{i}az-Rubio}%
\affiliation{%
Department of Electronics and Nanoengineering, Aalto University, P.~O.~Box~15500, FI-00076 Aalto, Finland
}%

\author{Junfei Li}
\affiliation{ 
Department of Electrical and Computer Engineering, Duke University, Durham, North Carolina 27708, USA
}%

\author{Steven A. Cummer}%
 \email{cummer@ee.duke.edu}
\affiliation{ 
Department of Electrical and Computer Engineering, Duke University, Durham, North Carolina 27708, USA
}%

\date{\today}

\begin{abstract}

We propose the design and measurement of an acoustic metasurface retroreflector that works at three discrete incident angles. An impedance model is developed such that for acoustic waves impinging at -60 degrees, the reflected wave is defined by the surface impedance of the metasurface, which is realized by a periodic grating. At 0 and 60 degrees, the retroreflection condition can be fulfilled by the diffraction of the surface. The thickness of the metasurface is about half of the operating wavelength and the retroreflector functions without parasitic diffraction associated with conventional gradient-index metasurfaces. Such highly efficient and compact retroreflectors open up possibilities in metamaterial-based acoustic sensing and communications.

\end{abstract}

\pacs{Valid PACS appear here}
\maketitle


Recent advances in acoustic metamaterials have revolutionized the manipulation of acoustic waves \cite{cummer2016controlling,ma2016acoustic,ge2017breaking}. By carefully engineering the subwavelength scatterers embedded inside the metamaterials, one is able to control the acoustic waves in a fashion not attainable with natural materials. The two-dimensional equivalent of metamaterials, i.e., metasurfaces, have attracted significant attention recently due to their thin feature and extraordinary capability of redirecting incident waves. Numerous applications have been proposed based on the concept of acoustic metasurfaces, such as wavefront modulation \cite{xie2014wavefront,tang2014anomalous}, sound absorption \cite{ma2014acoustic,cai2014ultrathin,li2016acoustic}, asymmetric transmission \cite{shen2016asymmetric,li2017tunable} and so on. In the scenario of reflection-based metasurfaces, the most extensively studied applications have been anomalous reflection \cite{li2013reflected}, negative reflection \cite{wang2016subwavelength,liu2017allangle}, carpet cloaking \cite{esfahlani2016acoustic} and more. In most of these cases, metasurfaces are considered single-channel systems where the desired response is obtained for a single direction of illumination or within a narrow range around the designed direction. 

Recently, a new concept of metasurfaces has been proposed that treats them as multichannel systems where different directions of illumination can be considered as channels \cite{asadchy2017flat}.  One promising possibility offered by multichannel metasurfaces is to perform the retroreflection functionality, i.e., incident waves from different directions can be reflected toward the impinging directions. By redirecting the incident energy into the original direction, such retroreflectors can find applications in free-space communications, remote sensing, and object detection and tracking. Literature on metasurface-based retroreflectors remains scarce in the field of acoustics \cite{fu2018compact}. In electromagnetics, retroreflectors have been proposed based on cat's eye geometries \cite{takatsuji1999whole} and Eaton lenses \cite{ma2009an}. These reftroreflectors, however, are bulky with sizes much larger than the operating wavelength. Metasurface-based retroreflectors have been proposed that stack two layers of metasurfaces \cite{arbabi2017planar}. Their efficiency is limited, however, due to the dual-layer configuration. Despite the deep accumulated knowledge on single-channel retroreflectors (or isolating mirrors) such as blazed gratings at optical frequencies \cite{bunkowski2006optical}, it was not until recently that a planar metasurface retroreflector was introduced in electromagnetics \cite{asadchy2017flat}. By tailoring the surface impedance profile, the metasurface can be designed to function as a retroreflector at multiple discrete angles. Such impedance-based flat retroreflectors avoid parasitic scattering and have high efficiency for multiple incident angles. 

In this paper, we investigate the possibilities of multichannel metasurface retroreflectors in acoustics. Based on the general surface impedance model \cite{asadchy2016perfect,diaz2017acoustic} and diffractive acoustics \cite{xie2014wavefront,wang2016subwavelength}, we design a metasurface retroreflector that functions at three specific angles, namely -60, 0, and 60 degrees. By enforcing impedance matching and reciprocity, the design in principle has 100\% efficiency and no parasitic scattering occurs at the surface. The metasurface retroreflector can have potential applications in acoustic communications, sensing and detection.

\begin{figure}
		\subfigure[]{\includegraphics[width=0.6\linewidth]{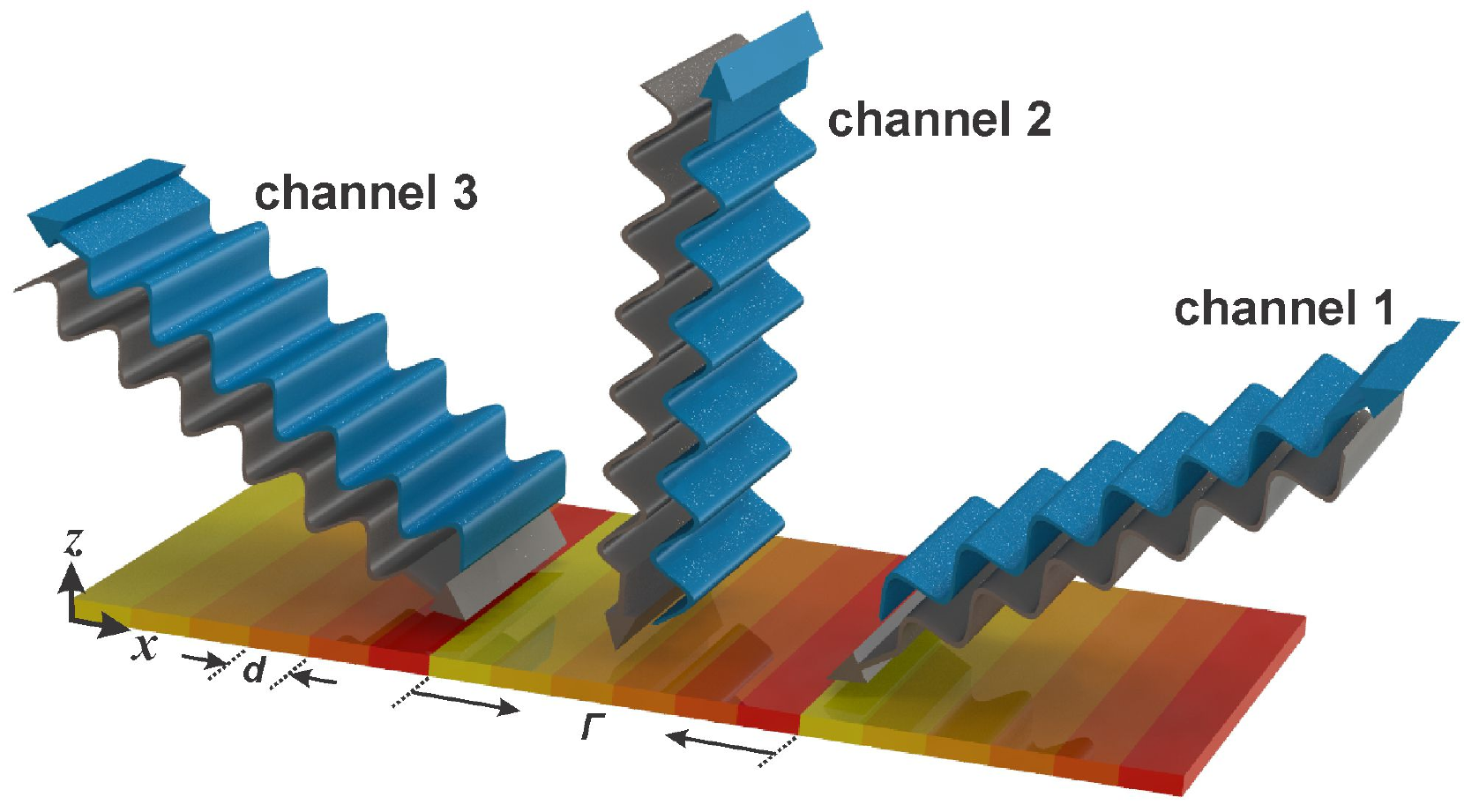}\label{fig:Fig1A}}
		\subfigure[]{\includegraphics[width=0.35\linewidth]{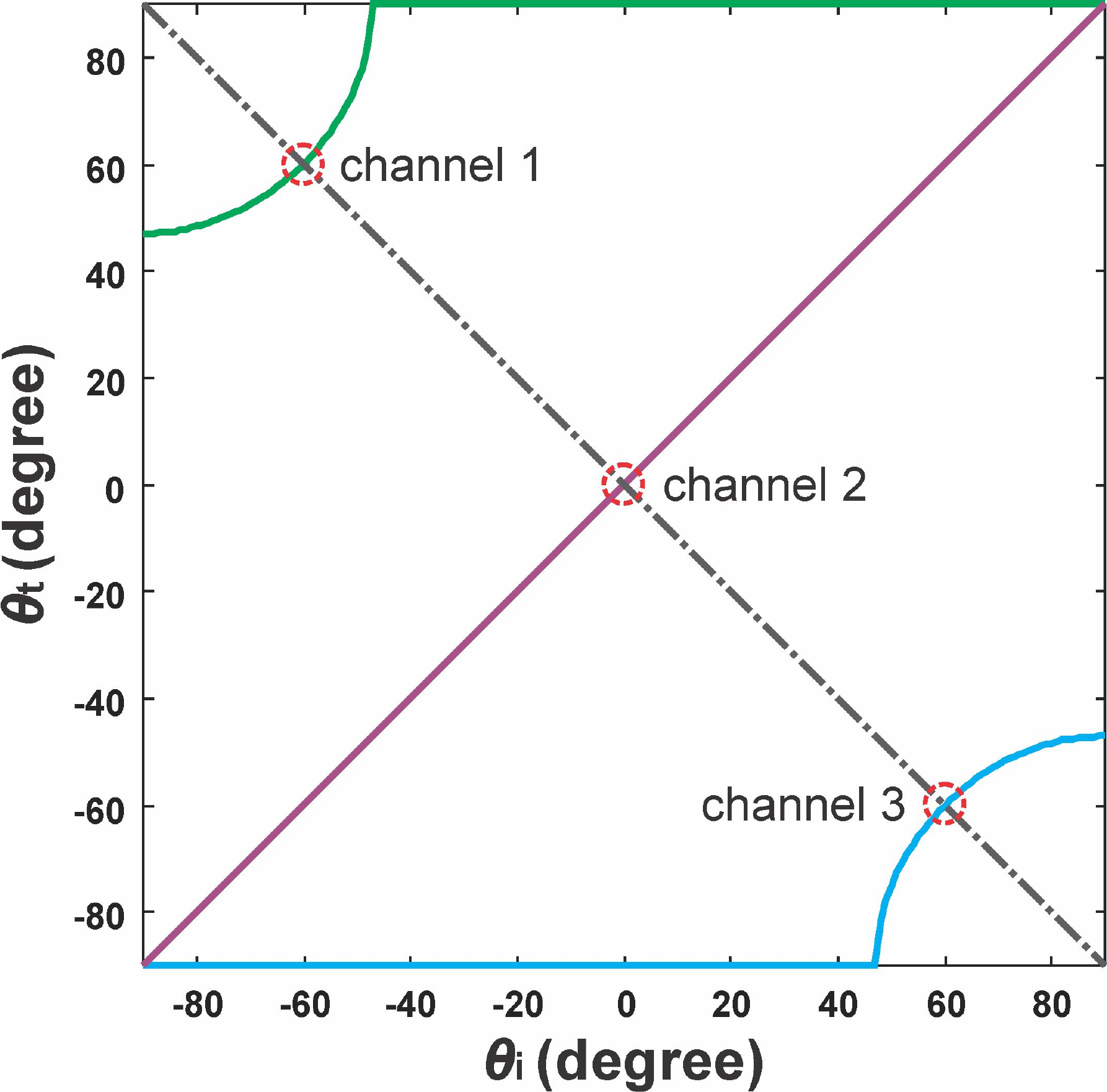}\label{fig:Fig1B}}
		\caption{(a) Illustration of the operation principle of the three-channel metasurface retroreflector. All the incident energy is reflected back toward the original direction without parasitic scattering. The gray arrows represent the incident waves and the blue arrows represent the reflected waves. (b) Diffraction analysis of the metasurface. Blue, purple and green curves represent the -1st, 0th, and +1st diffraction orders. Dotted gray curve shows the requirement for an ideal retroreflector. The three channels are marked by red circles.}
\label{fig:Fig1}
\end{figure}

The schematic illustration of the metasurface retroreflector is shown in Fig. \ref{fig:Fig1A}, where a flat metasurface extends in the $x$ direction. The width of an individual unit cell and the periodicity of the metasurface are $d$ and $\Gamma$, respectively. When the incident acoustic wave reaches the metasurface, it will be reflected toward the original incidence direction without any parasitic scattering. To ensure perfect efficiency, we require that all the incident energy is reflected with the pressure amplitude preserved. The incident and reflected acoustic pressure fields can therefore be written as:
\begin{eqnarray}
p_i(x,z)={p_0}e^{-j{k_x}x}e^{j{k_z}z}\\
p_r(x,z)={p_0}e^{j{k_x}x}e^{-j{k_z}z}
\label{eq:acousticfields}
\end{eqnarray}
where $p_0$ is the amplitude of the incident wave, $k_x=k\sin\theta _i$ and $k_z=k\cos\theta _i$ are the $x$ and $z$ components of the wave number, with $\theta _i$ being the incident angle. $k={\omega}/c$ is the free-space wave number, $\omega$ is the angular frequency and $c$ is the sound speed in air. The surface impedance relating the incident and reflected waves can be found by:
\begin{equation}
Z_s=\frac{p_{tot}(x,0)}{-\hat{n}\cdot\vec{v}_{tot}(x,0)}
\label{eq:theoryimpedance}
\end{equation}
where $\hat{n}$ is the unit vector normal to the metasurface, $p_{tot}$ and $\vec{v}_{tot}$ denote the total acoustic pressure and velocity. By recognizing that $\vec{v}=j\nabla p/\omega \rho$, the surface impedance can be obtained as:
\begin{equation}
Z_s=j\frac{Z_0}{\cos{\theta _i}}\cot({k_i \sin{\theta _i}x})
\label{eq:realimpedance}
\end{equation}

Without loss of generality, we choose as an example $\theta _i=-60^\circ$ as channel 1. The behavior of other channels ($0^\circ$ and $60^\circ$) can be analyzed as following. From Eq. (\ref{eq:realimpedance}), the periodicity of the surface impedance is $\Gamma=\lambda /(2\sin {\theta _i})$, and the wave vectors of the diffracted waves can therefore have the following tangential components:
\begin{equation}
k_{xn}=k_x+2\pi n/\Gamma, n \in \mathbb{Z}
\label{eq:diffraction}
\end{equation}
The corresponding reflection angles of different diffraction orders at a given angle of incidence can be found from $\theta_{rn}=\sin ^{-1}{k_{xn}/k}$. Figure \ref{fig:Fig1B} shows the reflection angle as a function of the incident angle for different diffraction orders. It can be seen that since the periodicity $\Gamma$ is small, the only allowed diffraction orders are $n=-1,0,1$. At channel 2, i.e., $\theta _i=0^\circ$, only one diffraction order of $n=0$ represents a propagation mode and the channel is isolated from the other two diffraction orders. Therefore, at channel 2, the condition for retroreflection, i.e., $\theta _i=-\theta _r$ is automatically satisfied. At channel 3, although $n=0$ diffraction will result in a propagating mode with reflection angle $\theta_r=60^\circ$, it is prohibited by reciprocity. This is because for the reciprocal and passive system considered here, coupling from channel 3 to channel 1 is exactly the same as from channel 1 to channel 3. The acoustic waves at channel 3 will hence experience a $n=-1$ diffraction, with the reflected angle being $\theta_r=-60^\circ$. The retroreflection condition is thus fulfilled at all three channels by engineering the surface impedance and diffraction of the metasurface properly.

\begin{figure}
\includegraphics[width=0.85\linewidth]{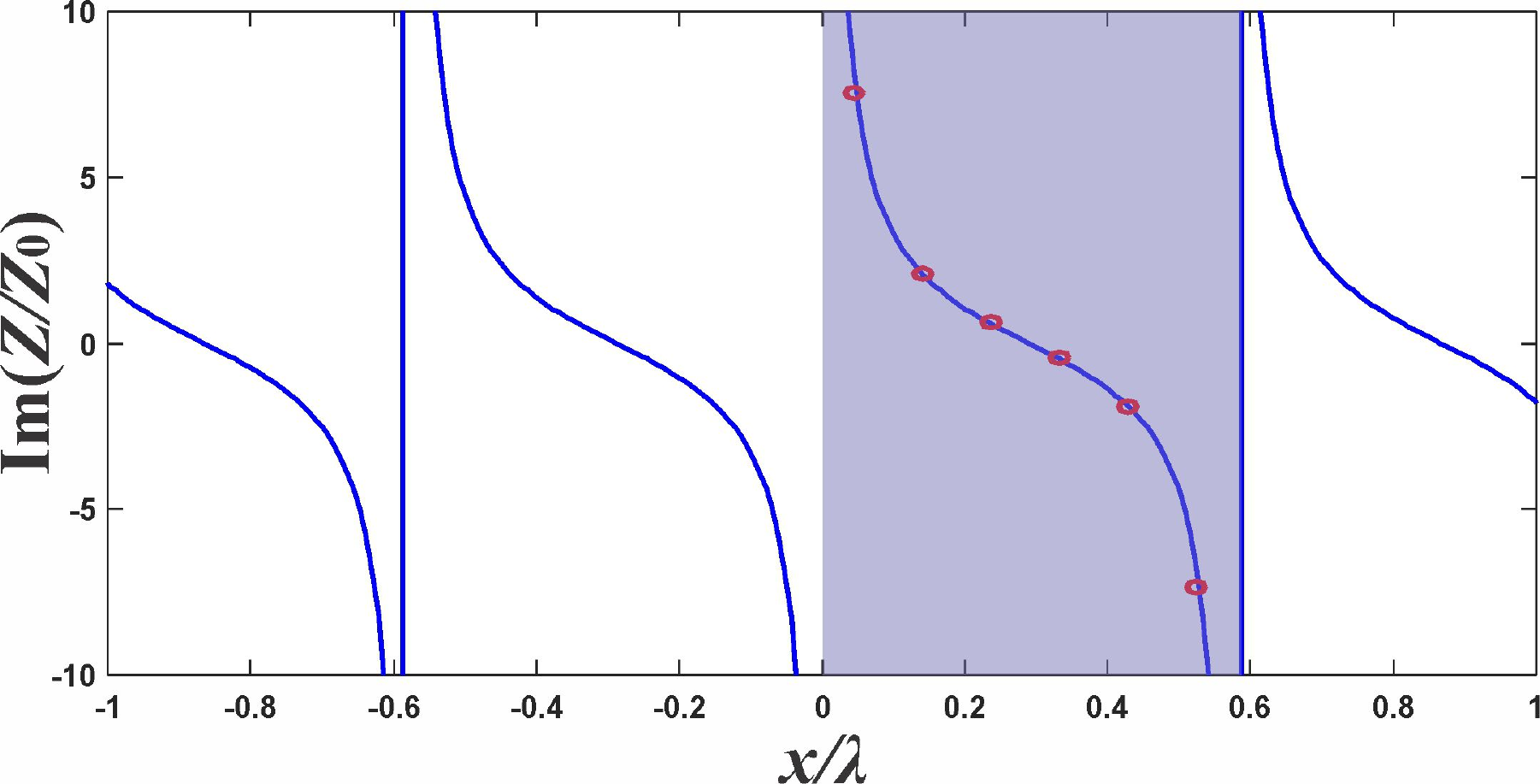}
\caption{Requirement of the surface impedance profile for channel 1. The shadow region denotes a period and the red circles mark the discretized impedance by the unit cells of the metasurface.}
\label{fig:Fig2}
\end{figure}

We now design the metasurface retroreflector using closed-end tubes \cite{kinsler1999acoustic}. For the selected incident angle of -60 degrees and operation frequency of 3000 Hz, it can be calculated from Eq. (\ref{eq:realimpedance}) that the periodicity of the metasurface is $\Gamma=6.6$ cm. To implement the required surface impedance profile, each period is discretized into six unit cells, with the width being $d=1.1$ cm. The height of each tube $l_n$ can be obtained through the relation $Z_{tube}=-jZ_0\cot (kl_n)$, where $n=1-6$ denotes the unit cell number. The first unit cell is positioned at $x=0.5d$ to avoid the extreme points, and the discretized impedance of the tubes is marked by circles in Fig. \ref{fig:Fig2}. The heights of the unit cells are 0.53 cm, 1.13 cm, 2.25 cm, 4.04 cm, 5.16 cm and 5.76 cm, respectively, all less than half of the operation wavelength at 3000 Hz. Since the closed-end tubes do not contain small channels or resonators, it is expected that the dissipation loss associated with this structure is low. In principle, other structures can also be used given that the impedance profile is satisfied. For example, space-coiling architectures \cite{xie2014wavefront,xie2013tapered}, helical structures\cite{zhu2016implementation}, shunted Helmholtz resonators \cite{li2015metascreen}, etc. can also be adopted, and the overall thickness of the metasurface may be further reduced. 

\begin{figure}
\includegraphics[width=0.85\linewidth]{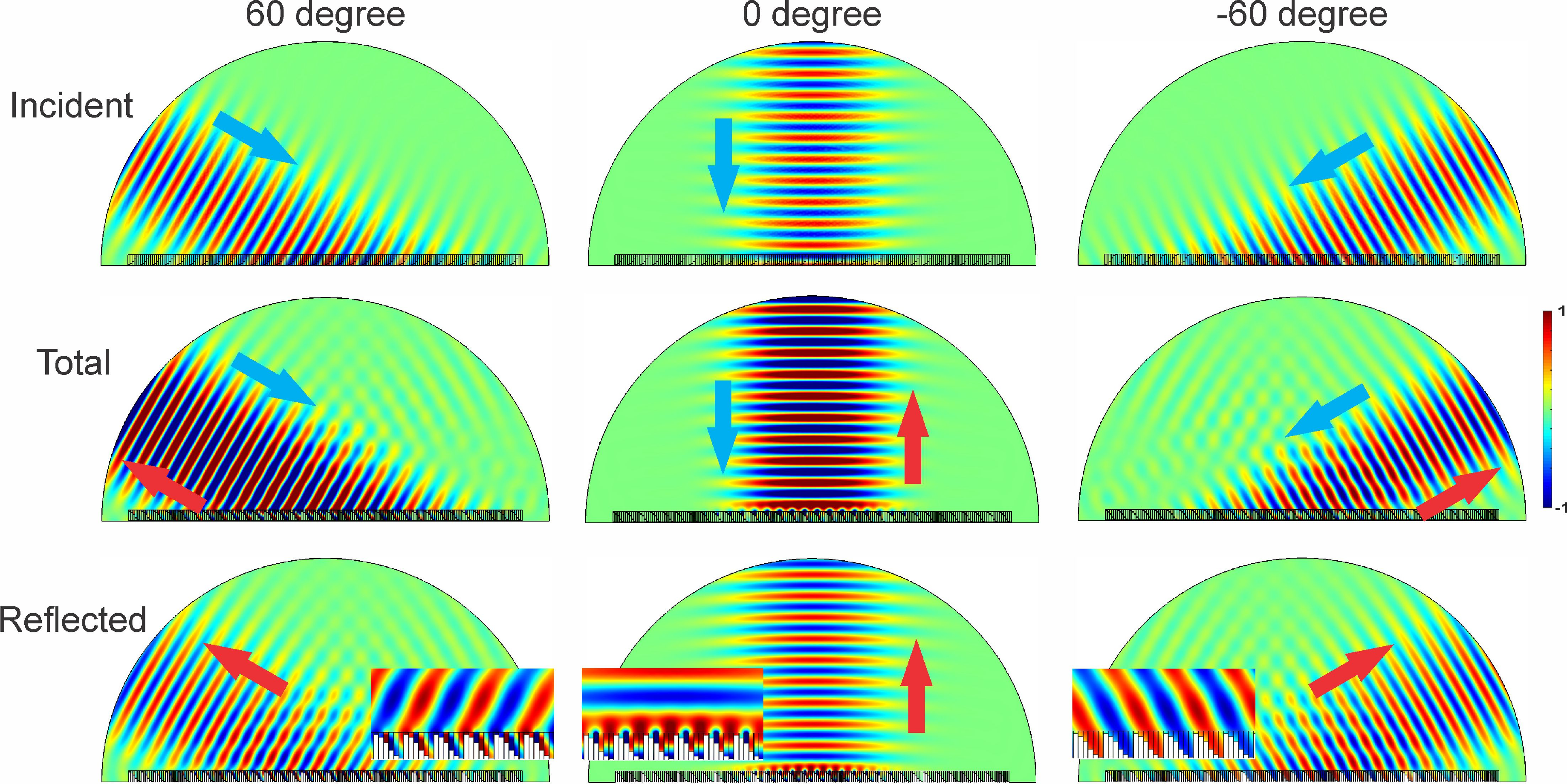}
\caption{Numerical simulations of the metasurface retroreflector. The top, middle and bottom panels show the incident, total and reflected acoustic fields. The blue and red arrows denote the incident and reflected beams, respectively. The inset shows the zoom-in field near the metasurface.}
\label{fig:Fig3}
\end{figure}

Numerical simulations based on the finite element package COMSOL Multiphysics are then performed to validate the performance of the metasurface. The tubes are made of acrylonitrile butadiene styrene (ABS) plastic with density 1230 kg/m$^3$ and sound speed 2230 m/s. The walls of the tubes are 1 mm thick and are assumed to be acoustically rigid since their impedance is much higher than that of the air. The corresponding acoustic fields are shown in Fig. \ref{fig:Fig3}, where a  spatially modulated Gaussian beam illuminates the metasurface at the designed angles. The reflected waves interact with the incident beams and an interference pattern can be observed. By subtracting the incident waves from the total acoustic fields, the reflected fields can be obtained. It can be clearly seen that the incident energy is redirected into the original direction and the amplitude is preserved. The small imperfectness of the reflected fields (e.g., parasitic scatterings) can be caused by the discretization of the impedance profile and the non-negligible thickness of the tube walls. Interestingly, it can be observed that some additional evanescent components are excited near the metasurface at $60^\circ$ and normal incidence. This is because at these angles, the reflected fields are formed by the diffraction on the metasurface, as shown in Fig. \ref{fig:Fig1B}. However, the evanescent waves decay rapidly away from the metasurface and do not contribute to the far-field efficiency. In the $-60^\circ$ case the reflected field contains only one plane wave since it is defined by the surface impedance from Eq.(\ref{eq:realimpedance}). The excellent agreement between the theory and simulations indicates that the retroreflector can reach almost 100\% efficiency by rerouting all the energy into the desired direction.

\begin{figure}
\includegraphics[width=0.85\linewidth]{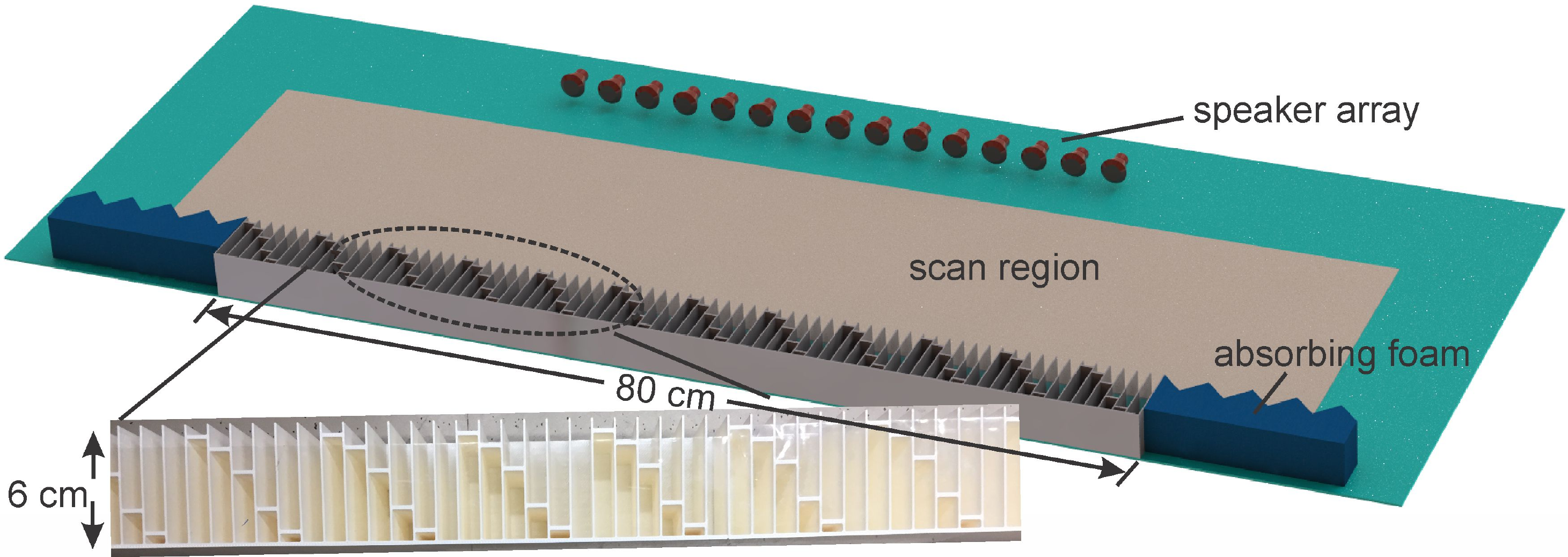}
\caption{Experiment setup of the metasurface retroreflector. The starting line of the scan region is 1 cm away from the exiting surface (tube openings) of the metasurface. The inset shows a portion of the fabricated sample.}
\label{fig:Fig4}
\end{figure}

\begin{figure}
\includegraphics[width=1\linewidth]{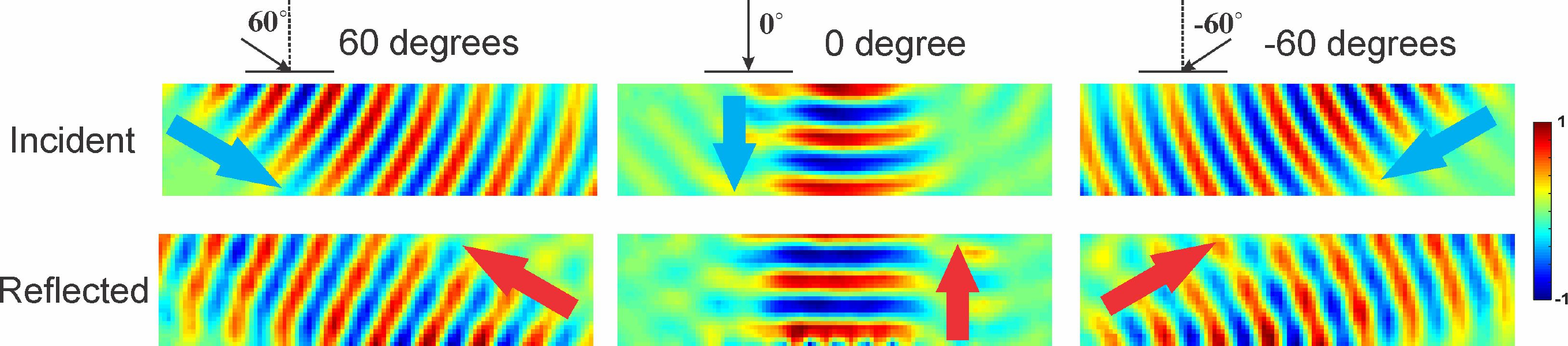}
\caption{Measured acoustic pressure fields at 3000 Hz. Top panels show the incident acoustic fields, bottom panels show the reflected acoustic fields.}
\label{fig:Fig5}
\end{figure}

The metasurface retroreflector is further validated with a 3D printing prototype. A portion of the fabricated sample is shown in the inset of Fig. \ref{fig:Fig4}. Each unit cell is connected to a back cavity so that the resulting metasurface has a flat surface. The back cavities are sealed with rigid printing material and are assumed to have no interaction with the incoming acoustic waves. The overall size of the sample is 80 cm by 6 cm, and the thickness is about half of the operation wavelength. The sample is situated in a 2D waveguide of 4 cm height to ensure that only the fundamental mode can propagate inside. A loudspeaker array consisting of 28 individual transducers is placed in front of the sample for the generation of incident Gaussian beams with 20 cm width. Absorbing foams are used to reduce the reflections on the edges. The acoustic field distribution inside the scan region is captured by a moving microphone with the step of 2 cm. The size of the scan area is 100 cm by 24 cm. The loudspeaker array is moved so that the beam fully illuminates the sample at different incident angles. Figure \ref{fig:Fig5} depicts the measured incident and reflected acoustic fields at different angles of incidence at 3000 Hz. Good agreement is observed between the numerical simulations and experiments, including the evanescent field pattern near the metasurface at normal incidence. The results confirm that the proposed metasurface can fully reflect the incident energy back toward its original direction without parasitic scatterings. 

\begin{figure}
\includegraphics[width=0.6\linewidth]{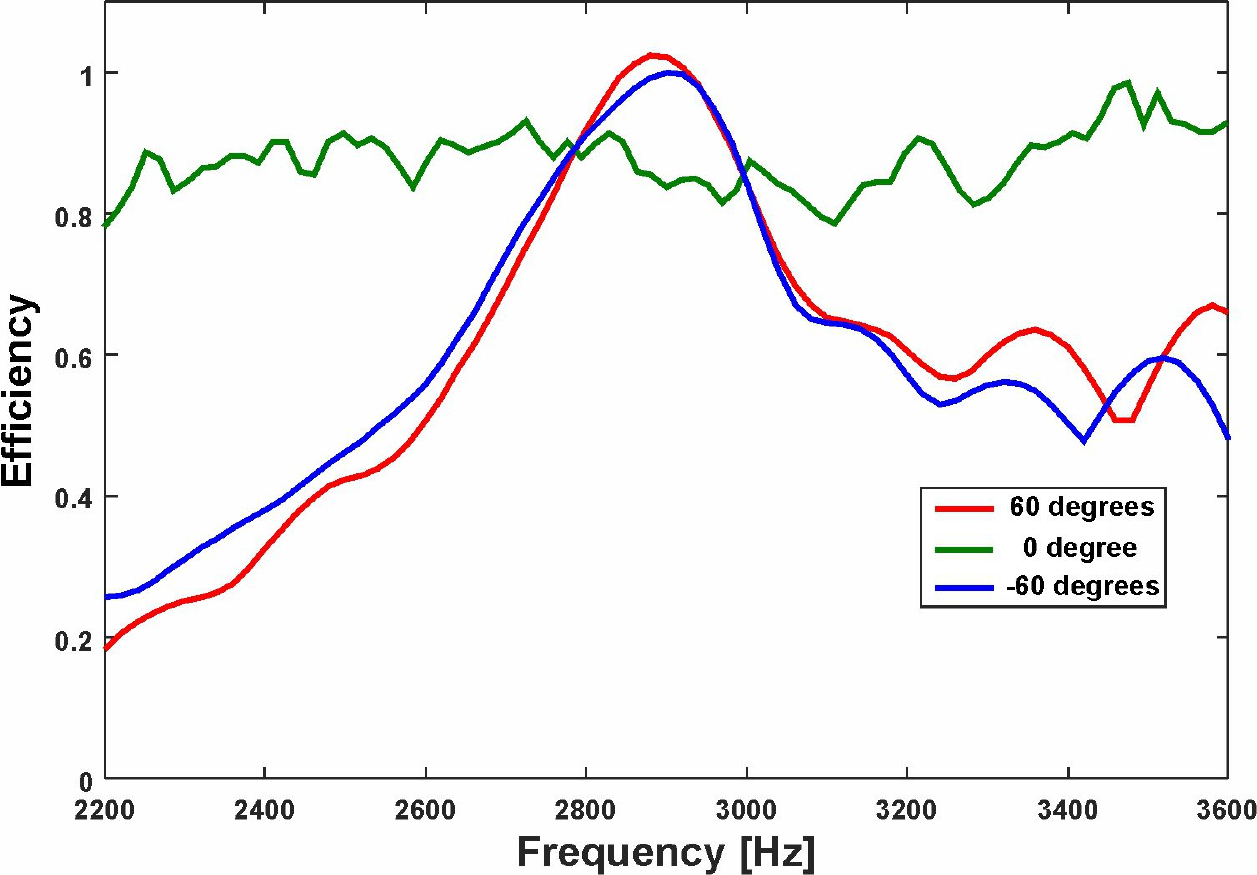}
\caption{Measured retroreflection efficiency of the metasurface at different angles of incidence.}
\label{fig:Fig6}
\end{figure}

To quantitatively study the performance of the metasurface retroreflector, the reflection efficiency is analyzed as a function of frequency. The efficiency is computed by dividing the reflected energy toward the desired direction by the incident energy. Specifically, the incident and reflected energy are calculated by performing a spatial Fourier transform along the line exiting the metasurface. The overall measured efficiency is shown in Fig. \ref{fig:Fig6}, with the maximum efficiency close to 100\%. The result well confirms the high efficiency and absence of parasitic scattering predicted by the theory. The peak frequency is slightly shifted from 3000 Hz to around 2900 Hz, which may be a result of fabrication errors. The efficiency is above 60\% within a frequency band of about 500 Hz for the $60^\circ$ and $-60^\circ$ cases and gradually decreases off the center frequency. This can be explained by the dispersive nature of the closed-end tubes, as the required impedance profile cannot be preserved when the frequency is far off the designed frequency. Remarkably, the efficiency for normal incidence is generally above 80\% and is not dependent on the frequency of the incident waves. This is because the periodicity is small and channel 2 is isolated, as can be seen in Fig. \ref{fig:Fig1B}. No other propagating modes are allowed for normal incidence as long as the periodicity remains small compared to wavelength, and only specular reflection occurs at the interface of the metasurface. The maximum efficiency is also slightly lower near peak frequency for the $0^\circ$ case, which may be because the beam width is relatively small (Fig. \ref{fig:Fig5}) and the diffractions on the edges are more profound.

To conclude, we have designed a high-efficiency three channel acoustic metasurface retroreflector that operates at three specific angles. The response of the metasurface is dictated by the engineered surface impedance and diffraction. Such surface impedance-based design strategy can reduce unwanted scatterings and yield a high efficiency. The proposed retroreflector is verified both numerically and experimentally with a 3D printed prototype. Measurements show that the retroreflector reaches nearly 100\% efficiency around the designed frequency. The device features a flat geometry and subwavelength thickness, and can be conveniently integrated into different scales or modified to work at other frequencies. The retroreflector can also in principle be extended for 3D wave propagation, or modified to have other functionalities such as anomalous reflection \cite{diaz2017acoustic}. Such a compact, multi-channel and high efficient retroreflector is hoped to be useful in acoustic sensing and communications.     

\begin{acknowledgments}
This work was supported by the Multidisciplinary University Research Initiative grant from the Office of Naval Research (N00014-13-1-0631) and in part by the Academy of Finland (projects 13287894 and 13309421).
\end{acknowledgments}

\nocite{*}
\bibliography{aipsamp}

\end{document}